# Uso de la conectividad cerebral basada en EEG para el estudio de la dinámica cerebral en interfaces cerebro-computadora
## Using EEG-based brain connectivity for the study of brain dynamics in brain-computer interfaces

**J.A. Gaxiola-Tirado**

Programa de Doctorado en Tecnologías Industriales y de Telecomunicación
Brain-Machine Interface Systems Lab
Universidad Miguel Hernández de Elche

Correspondencia/Correspondance: jorgen.gaxiola@goumh.umh.es

## RESUMEN
El análisis de la conectividad cerebral tiene como objetivo comprender la aparición de redes funcionales en el cerebro. Esta información puede utilizarse en el proceso de análisis y clasificación de señales de electroencefalografía (EEG) en aplicaciones de interfaz cerebro-computadora (BCI: *Brain Computer Interfaces*). Estos sistemas proporcionan un canal de comunicación y control a personas con discapacidades motoras. En este artículo, se exponen cuatro estrategias para emplear la conectividad cerebral en un entorno de BCI, a fin de obtener una mejor comprensión de los mecanismos cerebrales subyacentes al ejecutar cierta tarea mental, con el principal objetivo de desarrollar un esquema orientado a la neuro-rehabilitación de la marcha en combinación con distintas neurotecnologías y exoesqueletos. Dicho esquema permitiría mejorar las estrategias existentes y/o diseñar nuevos enfoques de control, así como de rehabilitación.

**Palabras clave:** Conectividad Cerebral, Interfaz Cerebro-Computadora, Rehabilitación.

## ABSTRACT
The analysis of brain connectivity aims to understand the emergence of functional networks into the brain. This information can be used in the process of electroencephalographic (EEG) signal analysis and classification for a brain-computer interface (BCI). These systems provide an alternative channel of communication and control to people with motor impairments. In this article, four strategies for using the brain connectivity in a BCI environment as a tool to obtain a deeper understanding of the cerebral mechanisms are proposed, with the principal aim of developing a scheme oriented to neuro-rehabilitation of gait in combination with different neurotechnologies and exoskeletons. This scheme would allow improving current schemes and/or to design new control strategies, as well as rehabilitation approaches.

**Keywords:** Brain Connectivity, Brain Computer-Interface, Rehabilitation.

# INTRODUCCIÓN

Una interfaz cerebro-computadora (BCI: *Brain-Computer Interface*) es un dispositivo que permite la comunicación de una persona con su entorno a partir de su actividad cerebral sin la asistencia de nervios periféricos o actividad muscular [1]. Hasta el día de hoy, se han desarrollado diversas aplicaciones de BCI, en su mayoría diseñadas con el propósito de mejorar el nivel de vida de las personas con algún tipo de deficiencia motora [2]. En este contexto, las aplicaciones más actuales están enfocadas en reemplazar o restaurar funciones motrices, ya sea de manera asistencial o con un enfoque de rehabilitación. Una *BCI asistencial* tiene como objetivo sustituir la pérdida de funciones motrices, mientras que una *BCI para rehabilitación* busca facilitar la restauración de las funciones cerebrales y/o motoras, como es el caso de las orientadas a rehabilitación de accidentes cerebrovasculares [3].

Sin embargo, a pesar del potencial de las BCI desarrolladas, existen todavía varios desafíos que limitan su impacto social y en investigación, ya que una notable parte de los usuarios estimada entre el 15 y el 30%, no es capaz de controlar estos dispositivos, este fenómeno es a menudo llamado *"analfabetismo en BCI"* [4]. Se han explorado algunas posibles soluciones a este fenómeno, tales como nuevas modalidades de entrenamiento y nuevos esquemas de retroalimentación [5]. En estos estudios, se ha obtenido como conclusión que el análisis debe ser personalizado. Por lo tanto, es preciso establecer esquemas individualizados que estimulen la capacidad de generar patrones de actividad cerebral estables y distintos durante diferentes tareas mentales. Así, se hace necesario aplicar nuevos enfoques que permitan una mejor comprensión de los mecanismos, las dinámicas de actividad cerebral y su relación con ciertas tareas mentales, ya que la actividad cerebral involucrada en BCI implica cambios relacionados con la plasticidad neural. Una herramienta prometedora para estudiar este tipo de enfoques es el análisis de la conectividad cerebral.

Generalmente, el análisis de la conectividad cerebral se asocia a técnicas de neuroimagen como la resonancia magnética funcional (fMRI). Sin embargo, con el auge de las BCIs basadas en mediciones no invasivas, como la electroencefalografía (EEG) y la compatibilidad directa de esta técnica con otros métodos de imagen cerebral, se ha comenzado a utilizar como una herramienta en la neuroimagen. En [6], se presenta una revisión de métodos básicos que hacen del EEG una herramienta capaz de proporcionar información espacio-temporal con respecto a la función cerebral en aplicaciones clínicas y experimentales.

En este artículo, se repasan los conceptos de BCI, conectividad cerebral y se presentan cuatro estrategias de combinación de estas herramientas. Todo ello, con la finalidad de utilizar la BCI como un medio para conocer más acerca de la actividad cerebral desencadenada en el usuario al ejecutar cierta tarea mental. Finalmente, se propone un enfoque de rehabilitación de la marcha. Este enfoque, en combinación con distintas neurotecnologías, permitirá mejorar los esquemas de rehabilitación existentes o diseñar nuevas estrategias de control y rehabilitación.

# INTERFACES CEREBRO-COMPUTADORA

Los sistemas BCI miden señales neurofisiológicas, las procesan y producen comandos de control en sistemas computarizados que reflejan la intención del usuario, con la finalidad de restaurar funciones de comunicación y control asistencial para personas con discapacidades motoras. La generación de un comando de control exitoso en estos sistemas depende de un adecuado reconocimiento de patrones aplicado a las señales cerebrales, lo cual es comúnmente implementado mediante cinco etapas consecutivas que se pueden observar en la Figura 1.

Primeramente, es necesario llevar a cabo la *adquisición de la señal*, esta etapa consiste en registrar la actividad eléctrica cerebral relacionada a distintas tareas mentales. En sistemas BCI se ha extendido el uso del EEG ya que es una técnica no invasiva, de bajo costo y de fácil instrumentación. Una vez adquirida la señal se lleva a cabo su *procesamiento*, es decir, se transforma la señal registrada en comandos de control para un sistema computarizado con la finalidad inicial de controlar un dispositivo externo, este bloque se divide en tres etapas que actúan de forma secuencial: *preprocesamiento, extracción de características y clasificación.*

Durante la etapa de *preprocesamiento* se lleva a cabo una adecuación de la señal, esto es, eliminar artefactos debido a otros tipos de actividades como el movimiento ocular o el producido por la línea eléctrica. Una vez preprocesada la señal, se realiza la *extracción de características*, en esta etapa se traduce la señal cerebral en un conjunto de características correlacionadas con el fenómeno estudiado. Durante esta fase, se genera una representación reducida y significativa de la señal usualmente en forma de vector que es empleada en la etapa de clasificación. Finalmente, durante la clasificación todo sistema BCI trata de reconocer a través del análisis las características extraídas, a que actividad neuronal (también llamada "clase") corresponden, a fin de identificar la intención del usuario. Para esto, el sistema debe entrenarse previamente con un conjunto de datos representativos para cada clase. Una vez identificada la intención del usuario, se emite un comando de *control* vinculado con la actividad cerebral correspondiente, con el fin de controlar un dispositivo externo. Puede incluirse una etapa de *retroalimentación* al usuario para mejorar el desempeño del sistema BCI.

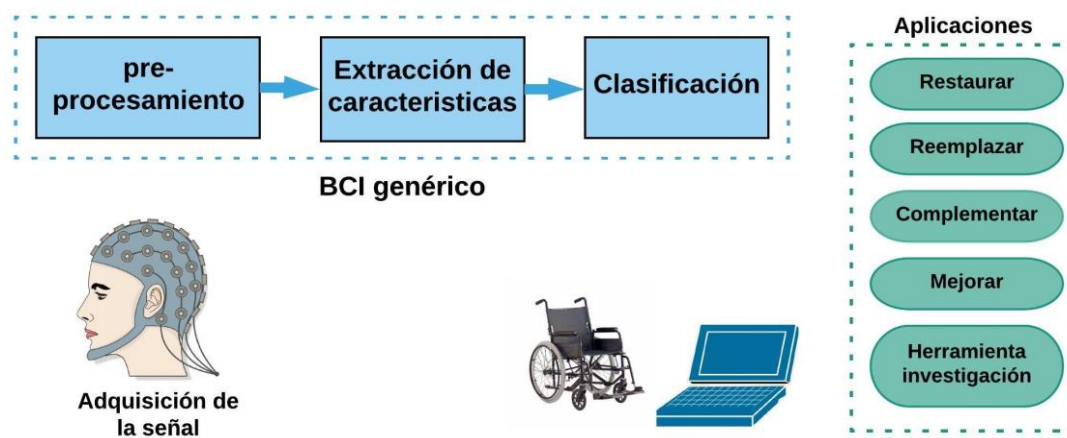

**Figura 1.** Etapas y aplicaciones de un BCI.

En el marco de las definiciones anteriores, la mayoría de las investigaciones de BCI se centran en la etapa de clasificación de señales, con un enfoque de control de algún dispositivo externo. Sin embargo, esto está empezando a cambiar, una definición más reciente de BCI es la siguiente: *"Un BCI es un sistema que mide la actividad del sistema nervioso central (SNC) y lo convierte en salida artificial que reemplaza, restaura, mejora o complementa la salida natural SNC y por lo tanto los cambios las interacciones en curso entre el SNC y su entorno externo o interno"* [7]. Los dos primeros tipos de aplicación de BCI, reemplazo o restauración funciones perdidas, son el foco de la investigación y desarrollo de las BCI más actuales. Al mismo tiempo, los otros tres tipos de aplicaciones están atrayendo cada vez más atención.

**Señales eléctricas utilizadas en BCI**

La actividad cerebral utilizada para aplicaciones BCI puede generarse de manera voluntaria, mediante la modulación de ciertos ritmos cerebrales, o de manera involuntaria, mediante la percepción de algún estímulo. En particular, las BCIs orientadas a la neuro-rehabilitación motora utilizan la actividad cerebral modulada voluntariamente. En este tipo de actividad, las oscilaciones cerebrales están siempre presentes y varían su amplitud dependiendo del estado mental del sujeto, de sus pensamientos o de determinadas acciones. Históricamente, los ritmos cerebrales se han dividido en cinco bandas de frecuencia: alfa (contiene al ritmo mu), beta, gamma, delta y theta. Los más utilizados para aplicaciones BCI son los ritmos sensoriomotores (SMR) mu y beta.

El ritmo mu, se encuentra en la banda de frecuencia de 8-13 Hz, y está relacionado con las funciones motoras. Su registro se realiza en la corteza motora en el lado opuesto al miembro en movimiento. Tiende a disminuir su amplitud cuando se realiza un movimiento, lo que se denomina desincronización (ERD: *Event-Related Desynchronization*). Se ha demostrado que se presenta un efecto similar de desincronización de este ritmo cuando se imagina el movimiento, característica que la hace útil como señal de control de un sistema BCI [8].

El ritmo beta, Se encuentra en la banda de frecuencia de 14-30Hz. Es común dividirlo en varias sub-bandas, una de ellas, la central (alrededor de los 18-26 Hz) se ve afectada por la imaginación o ejecución de movimientos. Se ha demostrado que este ritmo incrementa su amplitud después de la imaginación o ejecución de un movimiento, a lo que se denomina sincronización (ERS: *Event Related Synchronization*) [9].

Las BCIs basadas en los ritmos sensorimotores de EEG (MI-BCI), regularmente utilizan como paradigma la imaginación motora (MI: *Motor Imagery*) que se refiere al acto de imaginar una acción motriz específica sin llevar a cabo la ejecución de la misma. Se ha demostrado que, el MI puede modificar la actividad neuronal en las áreas sensoriomotoras primarias de una manera muy similar a la observada con movimientos ejecutados, y que estas pueden cambiar la conectividad funcional [10].

**Rendimiento de las BCI basadas en imaginación motora**

El rendimiento de las BCI basadas en MI, se mide regularmente con el porcentaje de acierto en la clasificación, lo cual depende en gran medida de una correcta extracción de características. Existen diferentes procedimientos para llevar a cabo este proceso como el filtrado espacial o el análisis espectral. En [11], pueden consultarse diferentes algoritmos usados en las etapas de extracción de características y clasificación. El uso de características basadas en tiempo y frecuencia han demostrado ser eficaces en el proceso de clasificación de tareas mentales. No obstante, proporcionan poca información sobre la actividad y las funciones cerebrales subyacentes. En la práctica una buena clasificación sin una explicación del fenómeno involucrado es poco útil.

La mayoría de los BCI utilizan el análisis de potencia espectral de señales adquiridas en canales únicos de EEG durante tareas de MI. Estas tareas involucran usualmente partes del cuerpo como las manos o los pies. De esta manera, los SMR modulan sus oscilaciones en función de la parte del cuerpo que se está imaginando. Particularmente, cuando el sujeto imagina el movimiento de alguna extremidad, los ritmos mu y beta se desincronizan sobre el área motora contralateral [12]. Es así, como el rendimiento de una BCI con este tipo de análisis es dependiente de la extracción de características basadas comúnmente en ERD/ERS, quedando espacialmente limitado a los lugares de registro de EEG en la corteza somatosensorial y motora.

En referencia a la observación anterior, se ha demostrado que durante el MI se activan y se comunican entre sí muchas partes del cerebro, específicamente el área motora suplementaria (SMA), área prefrontal, corteza premotora, cerebelo y los ganglios basales [13]. Por consiguiente, es necesario obtener características que proporcionen información sobre la relación entre las señales de adquiridas en diversos sensores de EEG. Para así, poder observar la interacción de áreas del cerebro separadas espacialmente. Lo anterior, se puede llevar a cabo a través del análisis de la conectividad funcional, la cual en los últimos años ha resultado ser una herramienta eficiente de la neurociencia.

# CONECTIVIDAD CEREBRAL

La comprensión y el modelado de la función cerebral no solo se basan en la correcta identificación de las regiones activas del cerebro, sino que también se consideran las interacciones funcionales entre los conjuntos neuronales distribuidos en diferentes regiones del cerebro. Estos conceptos se denominan en neurociencias como *segregación*, que se refiere a la activación de regiones especializadas del cerebro, lo que implica que una función cognitiva o motora se puede localizar en un área cortical específica, e *integración*, lo que está relacionado con la activación coordinada de un gran número de conjuntos neuronales distribuidos en diferentes áreas corticales que constituyen el procesamiento de las funciones cognitivas y las tareas motoras [14].

La integración de las áreas cerebrales se puede medir evaluando la conectividad cerebral. Esta herramienta permite analizar las interacciones entre distintas regiones del cerebro y se ha divido principalmente en tres tipos: estructural, efectiva y funcional; donde las dos últimas son relevantes para las aplicaciones BCI.

La *conectividad estructural* es difícil de definir debido a la escala microscópica de las neuronas. Se refiere a las conexiones sinápticas, proyecciones axonales y vías cerebrales dentro de ciertos grupos neuronales que ocurren dinámicamente [14]. Los tipos de conectividad funcional y efectiva generalmente se abordan mediante métodos matemáticos, basados en medidas de correlación o covarianza en series de tiempo multicanal registradas en diferentes ubicaciones. La *conectividad funcional* es la relación entre las regiones del cerebro que comparten propiedades funcionales. Se define como la correlación temporal (en términos de dependencia estadísticamente significativa) de la actividad neurofisiológica entre regiones del cerebro que podrían estar anatómicamente distantes. La *conectividad efectiva* puede verse como la unión de conectividad estructural y funcional. Se define como la influencia que un sistema neuronal ejerce sobre otro [14].

**Métodos computacionales para conectividad cerebral**

Los métodos de conectividad cerebral han sido clasificados inicialmente como basados en modelos y en datos. Los métodos basados en modelos necesitan estimaciones a priori sobre los mecanismos de interacción. Mientras que las técnicas basadas en datos no suponen ningún modelo subyacente específico o conocimiento previo sobre relaciones espaciales o temporales. Estos métodos se pueden utilizar para evaluar la conectividad cuando no se dispone de conocimiento estructural a priori, lo que los hace de utilidad en análisis BCI.

En el caso de un conjunto de señales de EEG registradas simultáneamente mediante distintos canales, éstas pueden ser vistas como un conjunto de datos multivariado. En este contexto, las métricas pueden ser clasificadas en bivariadas (se consideran dos canales a la vez) y multivariadas (considerando más de dos canales). En el caso de un sistema multivariado, son posibles diferentes patrones de conectividad. Cuando se utilizan medidas de conectividad bivariadas puede ser prácticamente imposible encontrar el patrón correcto de propagación.

En lo que respecta al análisis de conectividad, la herramienta de análisis más difundida es la correlación entre pares de señales. La principal dificultad de las métricas de correlación es que solo permiten identificar interacciones entre señales, pero no la direccionalidad de la interacción. Un concepto denominado Causalidad de Granger introduce la posibilidad de detectar la direccionalidad del flujo entre un conjunto multivariado de señales. Esto puede interpretarse como una indicación de cómo fluye la información a partir de una variable observada a otra, lo que conduce al término *interacción dirigida*. Una métrica que permite conocer las relaciones causales directas en el dominio de la frecuencia de la relación multivariable de señales de EEG es la Coherencia Parcialmente Dirigida (PDC).

Entre los métodos más representativos para estimar la conectividad cerebral se encuentran: el Modelado Causal Dinámico (DCM), la Magnitud Cuadrada de la Coherencia (MSC), Coherencia (COH), Valor de bloqueo de fase (PLV), Sincronización Generalizada (GS), Causalidad de Granger (GC) y la PDC. Las cuales, permiten el análisis de interacciones corticales complejas desde diferentes perspectivas novedosas [15]. En la Tabla 1, se resumen las principales características de los métodos mencionados.

| Característica | DCM | MSC | COH | PLV | GS | GC | PDC |
|---|---|---|---|---|---|---|---|
| Lineal | X | X | X | | | | |
| No lineal | | | | X | X | | |
| Basado en modelo | X | | | | | | |
| Basado en datos | | X | X | X | X | X | X |
| Causalidad | X | | | | X | | X |
| Multivariado | X | | | | X | | X |
| Conectividad funcional | | X | X | X | X | X | X |
| Conectividad efectiva | X | | | | | X | X |

**Tabla 1.** Comparación de métodos representativos para estimar la conectividad cerebral. Adaptado de [18].

## Uso de la Conectividad cerebral en esquemas BCI

En esta sección se realiza una revisión de artículos científicos publicados en los últimos cinco años, que contemplan el análisis de la conectividad cerebral basada en EEG, utilizando MI y enfocados al control de un BCI. Esta búsqueda, se realizó en los entornos de IEEE Xplore y ScienceDirect, bajo la ecuación de búsqueda: *(("brain computer interface" OR "BCI" OR "BMI") AND ("brain connectivity") AND "EEG")*. Se excluyeron investigaciones que incluían combinaciones de métodos de adquisición de señales y sujetos con patologías. Se encontraron un total de 27 contribuciones, 20 de publicadas en memorias de congresos internacionales y 7 en artículos de revistas (hasta el 26 de diciembre del 2018). En la Tabla 2, se resumen aspectos generales de los artículos de revista encontrados.

| Estudio | Paradigma | Método de conectividad | Resultados principales |
|---|---|---|---|
| Athanasiou et al., 2012. [16] | MI de la mano y pie | Función de trasferencia directa (DTF) | Se pudo discriminar redes de conectividad de imaginación motora (MI: motor imagery) de mano y pie en el 70% de los sujetos. |
| Daly et al., 2012 [17] | MI del dedo índice de la mano izquierda y derecha | EMDPL (Empirical Mode Decomposition Phase Locking) y coeficiente de agrupamiento | Porcentajes de clasificación mayores al 70% y siempre mayores en comparación al enfoque de potencia de espectral. |
| Krusienski et al., 2012 [18] | MI de manos | potencia espectral, magnitud de la coherencia al cuadrado (MSC) y PLV | La potencia espectral produjo una clasificación al menos tan buena como PLV, coherencia o cualquier combinación posible de estas medidas. |
| Billinger et al., 2013 [19] | MI de pies y manos | Transformada rápida de fourier (FFT) y medidas de conectividad basadas en vectores autorregresivos (VAR) | Mostraron que la clasificación de un solo ensayo MI es posible con medidas de conectividad extraídas de modelos VAR, y que un BCI podría potencialmente utilizar tales medidas. |
| Salazar varas et al., 2015 [20] | MI Pie y mano derechos | Coherencia ordinaria | El método propuesto logra buenos índices de clasificación de hasta 89% con un número reducido de sensores. |
| Gonuguntla et al., 2016 [21] | MI de mano izquierda y derecha | Valor de bloqueo de fase (phase-locking value, PLV) | El enfoque propuesto produjo una mejor clasificación en comparación con los enfoques basados en el análisis espectral. |
| Gaxiola et al., 2018 [22] | Mi Pie derecho y mano derecha. | Coherencia parcialmente dirigida (PDC) | Se propone un método de análisis de la conectividad cerebral direccional con aplicación a BCI. El método realizado puede ayudar a elucidar el porqué de las bajas tasas de clasificación para ciertos sujetos. |

**Tabla 2.** Estudios que utilizan la conectividad cerebral basada en EEG para aplicaciones de BCI

**Estrategias de uso de la Conectividad cerebral en BCI**

En esta sección se exponen cuatro estrategias que engloban la utilidad del análisis de la conectividad cerebral en esquemas de BCI. Estas estrategias, se pueden clasificar en esquemas directos e indirectos. Un esquema indirecto consiste en utilizar la BCI convencional y, posteriormente complementarla con el análisis de conectividad con la finalidad de correlacionar los resultados con la tarea mental en estudio (estrategia 2 y 4). Por otro lado, un esquema directo consiste en diseñar la BCI con base en el análisis de la conectividad (estrategia 1 y 3). Cabe mencionar que se pueden desarrollar nuevos esquemas de optimización de basados en la combinación de varias estrategias. Las estrategias se mencionan a continuación:

**Estrategia 1:** el análisis de conectividad cerebral se puede utilizar en la etapa de extracción de características con la finalidad de inferir los patrones de conectividad propios a cada tarea en estudio. Estos patrones, pueden ser tratados como el vector de características para llevar a cabo la clasificación y posteriormente enviar el comando al dispositivo externo como se puede observar en la Figura 2a.

**Estrategia 2:** una posible manera de medir la efectividad del entrenamiento es por medio de los cambios en la conectividad cerebral que ocurren durante en el transcurso de las sesiones de operación del sistema de BCI como se muestra en la Figura 3b. Esta estrategia es de utilidad en estudios de *analfabetismo BCI* y como una manera de cuantificar plasticidad neural durante el entrenamiento BCI.

**Estrategia 3:** esta estrategia consiste en utilizar la conectividad cerebral como medio de selección de las áreas cerebrales óptimas para obtener las señales (Figura

3c). Esto es, elegir aquellos electrodos que contengan características más informativas sobre la tarea mental en estudio. Esta estrategia puede ir de la mano con la estrategia 1, para esto es necesario llevar a cabo un análisis previo de caracterización de la conectividad cerebral.

**Estrategia 4:** las estrategias anteriores están enfocadas en adquirir características informativas para tener una comprensión más profunda del proceso de aprendizaje en una BCI. Un enfoque alternativo es recurrir a métodos para incrementar la plasticidad cerebral como exoesqueletos y la estimulación transcraneal por corriente directa (tDCS) para mejorar el rendimiento de una BCI. Entonces, el análisis de conectividad cerebral es útil para detectar sitios de estimulación o para analizar los cambios cerebrales presentados al utilizar alguna neurotecnología y/o un exoesqueleto (Figura 3d).

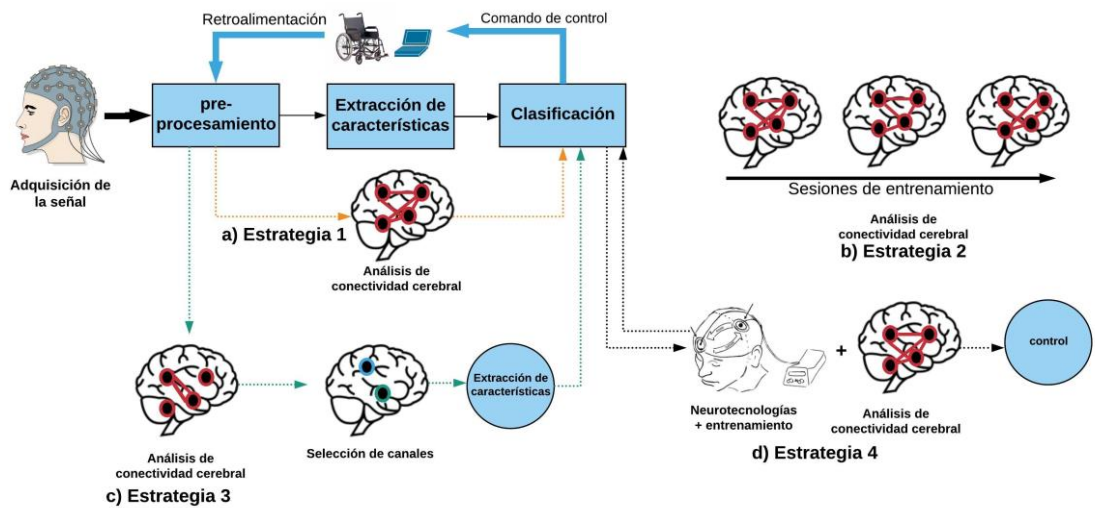

**Figura 2.** Estrategias de incorporación del análisis de la conectividad cerebral en esquemas de BCI.

**Combinación de estrategias para neuro-rehabilitación**

En los últimos años, se han desarrollado neurotecnologías para mejorar la calidad de vida de personas con deficiencias motrices, como BCIs o exoesqueletos robóticos. En este proyecto de doctorado se contempla realizar una estrategia para la rehabilitación de la marcha basada en la combinación de neurotecnologías, considerando el estudio de la conectividad cerebral como medio de conocer los procesos cerebrales desencadenados en esquemas BCI.

Recientemente, en el grupo de investigación BMI System Lab de la universidad Miguel Hernández de Elche, se ha diseñado una estrategia para la rehabilitación de la marcha, que incorpora la aplicación la estimulación transcraneal con corriente directa (tDCS) y un exoesqueleto de miembro inferior en un esquema de BCI [23]. En este caso, la BCI fue diseñada para detectar dos estados: relajación y MI de marcha; se basó en encontrar la máxima diferencia de potencia entre los dos estados mentales para cada electrodo de EEG. Se demostró que, en términos de potencia, el montaje aumentó la clasificación de MI de las extremidades inferiores. Además, se indicó que la utilización de un exoesqueleto en tiempo real tiene el potencial de mejorar el proceso de rehabilitación. Siguiendo esta línea de investigación, los objetivos contemplados en el presente proyecto de doctorado serán los siguientes:

**Objetivo 1:** añadir el análisis de conectividad cerebral a la estrategia para la rehabilitación de la marcha propuesta anteriormente en [23], como se indica en la

Figura 3b, esto con la finalidad de estudiar si analizando las regiones de conectividad cerebral es posible determinar si ha habido una mejora en la rehabilitación de la persona (se ha mejorado la neuroplasticidad). Para este fin, se utilizará el método basado en la PDC y enfocado a BCI propuesto en [22].

**Objetivo 2:** realizar la validación del exoesqueleto T-Flex [24] para la asistencia de la marcha, con el objetivo de diseñar un esquema de control de este exoesqueleto por medio de una BCI, el cual en combinación con el análisis de la conectividad cerebral, permitirá optimizar y mejorar este exoesqueleto. Además, este sistema podría permitir la práctica repetida de la actividad motora a rehabilitar, mediante la imaginación motora.

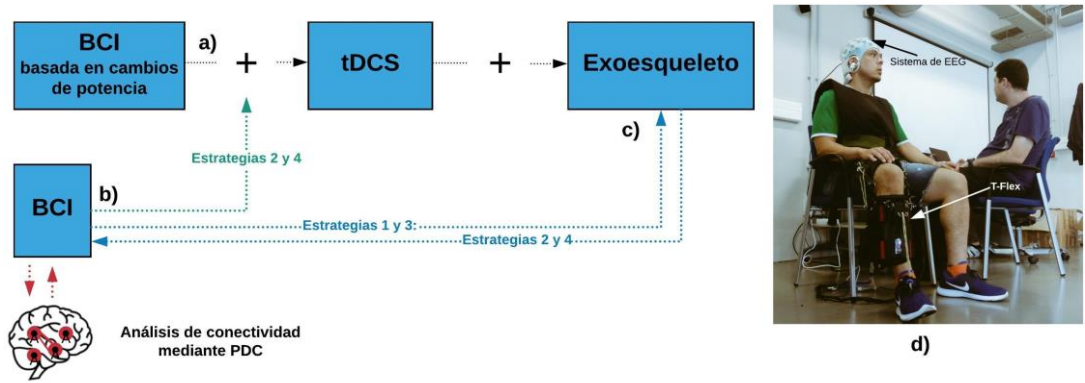

**Figura 3. Combinación de estrategias para neuro-rehabilitación.** **a)** Estrategia de rehabilitación de la marcha propuesta en [23]. **b)** Incorporación de la conectividad a la estrategia propuesta de rehabilitación. **c)** Esquema propuesto para la validación del exoesqueleto T-Flex mediante una BCI. **d)** Experimentación preliminar para el control de exoesqueleto T-Flex mediante una BCI basada en EEG.

## CONCLUSIÓN

La conectividad cerebral se ha convertido en una herramienta popular para comprender la aparición de redes funcionales en el cerebro. En este artículo, se proporciona una perspectiva sobre el como el estudio de la conectividad cerebral tiene el potencial de mejorar el rendimiento de diversos esquemas de BCI. Se presentan estrategias para utilizar la BCI no solo como un medio de control, si no como una herramienta para conocer más acerca de la actividad cerebral desencadenada en el usuario al ejecutar cierta tarea mental, para entonces, poder diseñar estrategias que de alguna manera coadyuven en el desempeño del usuario en diversas aplicaciones de BCI y neuro-rehabilitación.

En el ámbito de la rehabilitación, la conectividad cerebral se puede utilizar como características cerebrales complementarias en los BCI. Durante el presente proyecto de doctorado, se utilizará la combinación de estas dos herramientas y la unión de distintos métodos para incrementar la plasticidad cerebral, la cual se cree que es la responsable de los cambios y mejoras funcionales en la ejecución de tareas motoras durante la rehabilitación. En este contexto, se estudiará si analizando las regiones de conectividad cerebral se puede determinar si ha habido una mejora en la rehabilitación de la persona.

## REFERENCIAS